\documentclass{article}
\usepackage{amsfonts}

\usepackage{alife9}
\usepackage{graphicx}
\usepackage{amsmath}

\begin{document}

\title{Introduction to Random Boolean Networks}
\author{Carlos Gershenson \\
%EndAName
\mbox{} Centrum Leo Apostel, Vrije Universiteit Brussel. Krijgskundestraat
33 B-1160 Brussel, Belgium\\
cgershen@vub.ac.be \ http://homepages.vub.ac.be/\symbol{126}cgershen}
\maketitle

\begin{abstract}
The goal of this tutorial is to promote interest in the study of random
Boolean networks (RBNs). These can be very interesting models, since one
does not have to assume any functionality or particular connectivity of the
networks to study their generic properties. Like this, RBNs have been used
for exploring the configurations where life could emerge. The fact that RBNs
are a generalization of cellular automata makes their research a very
important topic.

The tutorial, intended for a broad audience, presents the state of the
art in RBNs, spanning over several lines of research carried out by
different groups. We focus on research done within artificial life,
as we cannot exhaust the abundant research done over the decades related
to RBNs.
\end{abstract}

\section{Introduction}

Random Boolean networks (RBNs) were originally developed by Stuart Kauffman 
 as a model of genetic regulatory networks \cite{Kauffman1969,Kauffman1993}. They
are also known as $N-K$ models, or Kauffman networks. RBNs are generic, because one does not assume any particular functionality
or connectivity of the nodes: these are generated randomly. This is a useful
approach if the specific structure and/or function of a system are very
complex and unknown. The generic properties found in the model can be then
applied to the particular system, in order to attempt to unveil its
mechanisms.

Mathematical and computational modelling of genetic regulatory networks
promises to uncover the fundamental principles of living systems
in an integrative and holistic manner. It also paves the way toward the
development of systematic approaches for effective therapeutic intervention
in disease \cite{ShmulevichEtAl2002}. Single-gene studies are very limited
for such intertwined networks. 

Even when there is a Boolean simplification, many systems can be studied with near-binary states.
This is because the behaviour of many systems is determined by thresholds, such as the ones determined by firing potentials of synapses in
neurons, or activation potentials of chemical reactions in metabolic
networks.

Furthermore, random Boolean networks have been applied and used as models in
many different areas, such as evolutionary theory, mathematics, sociology,
neural networks, robotics, and music generation.

In the next section, we will review the classic RBN model, its three characteristic phases (ordered, chaotic, and critical, some explorations of the model, alternatives, and extensions. Next we will review the effect of the updating scheme in RBNs: syncrhonous-asyncrhonous, determiistic-non-deterministic. We mention briefly some applications of RBNs, tools availabe for their study, and future lines of research.

\section{Classical Model}

Kauffman proposed the original RBN model, supporting the hypothesis that living organisms could be constructed from random elements, without the need of precisely programmed elements \cite{Kauffman1969}. Certain types of RBNs are very robust, and have many analogies to living organisms.

A RBN consists of $N$ nodes, which can take values of zero or one (Boolean).
The state (zero or one) of each node is determined by $K$ connections coming
from other (or the same) nodes. The connections are wired randomly, but
remain fixed during the dynamics of the network, i.e. ``quenched''. The way
in which nodes affect each other is not only determined by their
connections, but by logic functions, which are generated randomly, simply
using lookup tables for each node, which take the states of the connecting
nodes as inputs, and the state of the node as output. These also remain
fixed (quenched) during the dynamics of the network.

We can see that RBNs are a generalization of Boolean cellular automata (CA) 
\cite{vonNeumann1966,Wolfram1986,WuenscheLesser1992}, where the state of
each node is not affected necessarily by its neighbours, but potentially by
any node in the network. RBNs with $N=K$ are also called random maps.

The updating of the nodes in classic RBNs is \emph{synchronous}: the states
of nodes at time $t$+1 depend on the states of nodes at time $t$, so that
all nodes ``march in step''. We will see below that there can be drastic
differences if we change the updating scheme.

\begin{figure*}[t]
\begin{center}
\includegraphics[width=6.5in]{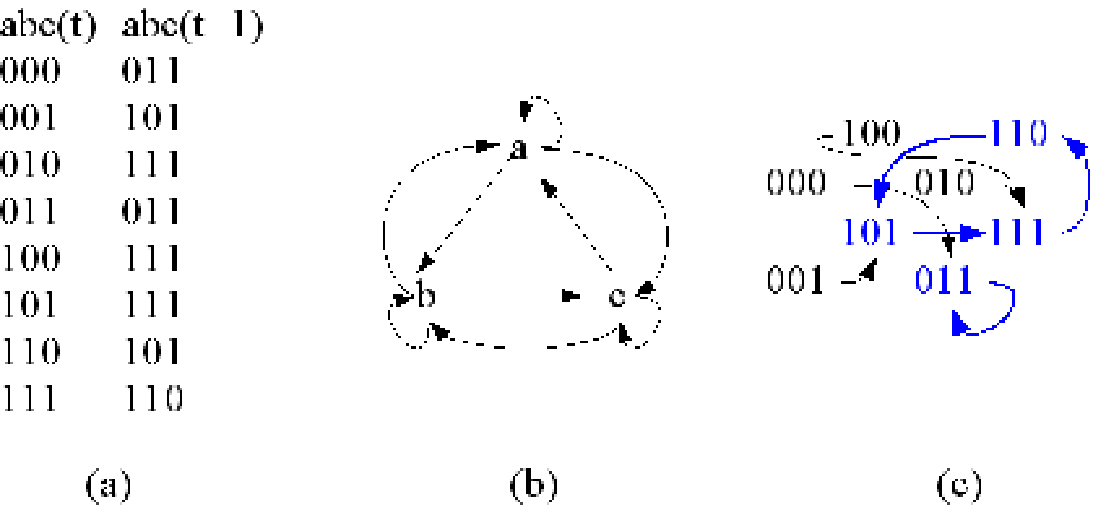}
\end{center}
\caption{a) Lookup table for the state transitions. b) Wiring diagram: in
this case, all nodes affect all nodes. c) State space diagram. There is one
point attractor (011), with one state flowing into it (000), and one cycle
attractor of period three (111$\rightarrow $110$\rightarrow $101), with
three states flowing into it (001, 010, 100).}
\label{VgrCRBN}
\end{figure*}

Usually, an initial random state is chosen for the RBN, and the dynamics
flow according to the updating functions and scheme. Since the state space
is finite ($2^{N}$), eventually a state will be repeated. Since the dynamics
are deterministic, this means that the network has reached an \emph{attractor%
}. If the attractor consists of one state, it is called a \emph{point}
attractor or steady state, whereas if it consists of two or more states, it
is called a \emph{cycle} attractor or state cycle. The set of states that
flow towards an attractor is called the attractor \emph{basin}. An example
of a RBN with $N=3$ and $K=3$ is shown in Figure \ref{VgrCRBN}.

If we try to imagine all possible networks, for each node there will be 2$%
^{2^{K}}$ possible functions. And each node has $N!/(N-K)!$ possible ordered
combinations for $K$ different links. Therefore all the possible networks for given $N$
and $K$ will be \cite{HarveyBossomaier1997}:

\begin{equation}
\left( \frac{2^{2^{K}}N!}{\left( N-K\right) !}\right) ^{N}
\end{equation}

Note that many of these will be equivalent, but nevertheless we can see that
the space of networks is immense. This makes things complicated for
statistical studies. There is not enough computational power to exhaust all
possible networks, so only ``representative'' properties can be studied. We
should also mention that there is a very high variance in the statistical
studies of RBNs.

However, general properties can be extracted from this huge universe of
possible networks.

\subsection{Order, Chaos, and the Edge}

In RBNs, as well as in many dynamical systems, three phases can be
distinguished: \emph{ordered}, $\emph{chaotic}$, and \emph{critical}. These
phases can be identified with different methods, since they have several
unique features.

If we plot the states of a network in a square lattice where the state of a
node depends topologically on its neighbours, and let the dynamics flow, we
can easily see which states change, and which ones are stable. In other
words, we can observe how much the network changes. We can colour changing
states with green, and static ones with red. In the ordered phase, we will
see that after we select an initial random state, initially many states are
changing (green), but quicky the dynamics stabilise, and most of the nodes
will be static (red). There will be only few green ``islands'', surrounded
by a red ``frozen sea''. In the chaotic regime, most of the states are
changing constantly, so we have a green sea of changes, typically with red
stable islands. The phase transition from the ordered to the chaotic regime,
also known as the ``edge of chaos'', occurs when the ordered green sea
breaks into green islands, and the red islands join and percolate through
the lattice \cite[pp. 166-167]{Kauffman2000}.

A second feature of these dynamical phases is related to ``sensitivity to
initial conditions'', ``damage spreading'', and ``robustness to
perturbations'' which are different ways of measuring the \emph{stability}
of a network. We can ``mutate'', ``damage'' or ``perturb'' a node of a RBN
by flipping its state. We can also change a connection between two nodes, or
in the lookup table of a node. Since nodes affect other nodes, we can
measure how much a random change affects the rest of the network. In other
words, we can measure how the damage \emph{spreads}. This can be done by
comparing the evolution of a ``normal'' network and a ``perturbed'' network.
In the ordered regime, usually the damage does not spread: a ``perturbed''
network ``returns'' to the same path of the ``normal'' network. This is
because changes cannot propagate from one green island to another. In the
chaotic phase, these small changes tend to propagate through the network,
making it highly sensitive to perturbations. This is because perturbations
can propagate through the percolating green sea \cite[pp. 168-170]
{Kauffman2000}. This ``butterfly effect'' is a common characteristic of
chaotic systems, where small perturbations can cause large consequences, and
systems are sensitive to their initial conditions. At the edge of chaos,
changes can propagate, but not necessarily through all the network.

A third feature is the convergence versus divergence of the trajectories in
state space of the network dynamics. In the ordered phase, similar states
tend to converge to the same state. In the chaotic regime, similar states
tend to diverge. At the edge of chaos, nearby states tend to lie on
trajectories that neither converge nor diverge in state space \cite[p. 171]
{Kauffman2000}.

Living systems, or computing systems, need certain stability to survive, or
to keep information; but also flexibility to explore their space of
possibilities. This has lead people to argue that life and computation occur
more naturally at the edge of chaos \cite{Langton1990}, or at the ordered
regime close to the edge of chaos \cite{Kauffman2000}. There could be
ordered or chaotic systems able to perform the same computations, but the
first ones would need more time, and the second more redundancy to cope with
their instabilities. Therefore, we can assume that an adequate balance of
order and chaos is more economic, thus preferable by natural selection.

\subsubsection{Phase Transitions in RBNs}

\begin{figure}[t]
\begin{center}
\includegraphics[width=3.2in]{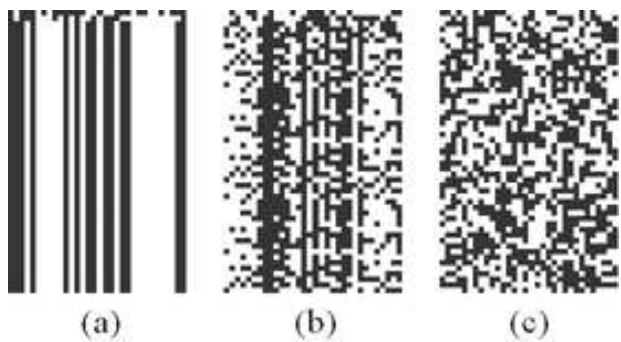}
\end{center}
\caption{Trajectories through state space of RBNs within different phases, $%
N=32$. A square represents the state of a node. Initial states at top, time flows downwards. a) ordered, $K=1$. b)
critical, $K=2$. c) chaotic, $K=5$}
\label{phasesDynExample}
\end{figure}

Very early in the studies of RBNs, people realized in simulations that the
networks with $K\leq 2$ were in the ordered regime, and networks with $K\geq
3$, were in the chaotic regime. In Figure \ref{phasesDynExample} we can
appreciate characteristic dynamics of RBNs in different phases.

We can identify phase transitions in RBNs in different ways. The main idea
is to measure the effect of perturbations, the sensitivity to initial
conditions, or damage spreading. This is analogous to Lyapunov exponents in
continuous dynamics.

The phase transitions can be statistically or analytically obtained. Derrida
and Pomeau were the first to determine analytically that the critical phase
(edge of chaos) was found when $K=2$ \cite{DerridaPomeau1986}. They also
introduced two generalizations of the classical model: one where they
consider nonhomogeneous networks ($K$ is not necessarily the same for all
nodes, so we use as a parameter the mean connectivity $\langle K\rangle $),
and another where the values of lookup tables have a probability $p$ of
being one (and thus $1-p$ of being zero).

The method they used, also known as the Derrida annealed approximation,
takes two random initial configurations, and measures their overlap. This
can be done with the normalized Hamming distance (\ref{eqHamming}). Then one
time step of the dynamics is computed, and the overlap is measured again.
Then, a new set of rules and connections is chosen at random. It can be
shown that this evolves in a one-dimensional map.

\begin{equation}
H\left( A,B\right) =\frac{1}{n}\sum_{i}^{n}|a_{i}-b_{i}|  \label{eqHamming}
\end{equation}
\qquad \qquad

\begin{figure}[t]
\begin{center}
\includegraphics[width=3.2in]{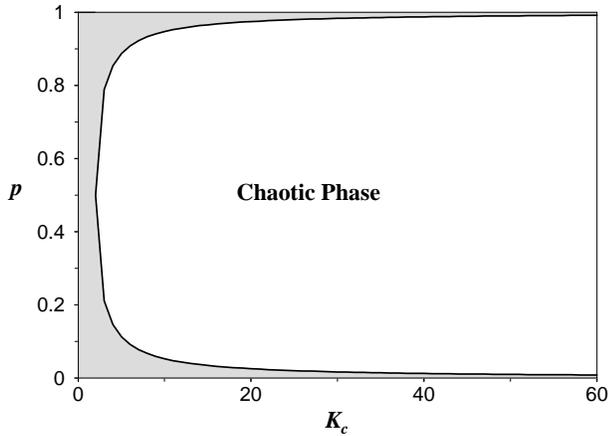}
\end{center}
\caption{Phase diagram for the classical model, reprinted from \protect\cite
{Aldana2003} with permission from Elsevier}
\label{Aldana-fig1}
\end{figure}

For $p=0.5$, the map converges to the stable point $H=0$ when $K<2$, meaning
that different states tend to converge (ordered dynamics). At $K=2$, this
point becomes unstable, meaning that for $K>2$, different states tend to
diverge (chaotic dynamics). It can be shown that the curve for critical $K$%
's depending on the value of $p$ follows the equation (\ref{eqDerrida}),
both for homogeneous and nonhomogeneous normal distributions of $K$. The
plot of this equation can be seen in Figure \ref{Aldana-fig1}.

\begin{equation}
\langle K\rangle =\frac{1}{2p\left( 1-p\right) }  \label{eqDerrida}
\end{equation}

Luque and Sol\'{e} made a simpler analytic determination of the phase
transitions in networks, where they study the damage spreading when single
nodes are perturbed \cite{LuqueSole1997}. They consider trees of nodes that
can affect the state of other nodes in time. As a node has more connections,
there will be an increase in the probability that a damage in a
single node (0$\rightarrow $1 or 1$\rightarrow $0) will percolate through
the network.

Let us focus only in one node $i$ at time $t$, and a node $j$ of the several 
$i$ can affect at time $t+1$. There is a probability $p$ that $j$ will be
one, and a damage in $i$ will modify $j$ towards one with probability $1-p$.
The complementary case is the same. Now, for $K$ nodes, we could expect that
at least one change will occur if $\langle K\rangle 2p(1-p)\geq 1$, which
leads to (\ref{eqDerrida}). This method can be also used for other types of
networks. Luque and Sol\'{e} later used the concept of Boolean derivative to
define Lyapunov exponents in RBNs \cite{LuqueSole2000}:

\begin{equation}
\lambda =\log \left[ 2p\left( 1-p\right) K\right]
\end{equation}

where $\lambda <0$ represents the ordered phase, $\lambda >0$ the chaotic
phase, and $\lambda =0$ the critical phase.

Statistical studies confirm these analytical results. However, it seems
that, \emph{in practice}, the size of the network can play a role in the
phase transitions \cite{Gershenson2004a,Gershenson2004b}. We have seen that
for large RBNs the phase transition (measured with the average of
differences of normalized Hamming distances at $t\rightarrow \infty $ and $%
t=0$, for minimum initial distances of $1/N$) is given for $K$ shifted
towards one, and for very small networks for $K$ shifted towards three. This
is probably because for large networks, there is a higher probability that a 
\emph{subnetwork} will be generating more noise than ``average'', thus
propagating damage.

In practice, there are very high standard deviations in RBN studies, which
are normal in this type of systems \cite{MitchellEtAl1993}. We can clearly
find (or design) ordered networks for $K\gg 3$ but on average, statistical
studies confirm the analytical ones. Therefore, when we generate a random
network, there are high probabilities that it will be in a specific regime
according to $K$ and $p$.

\subsection{Explorations of the Classical Model}

There have been several explorations of different properties of RBNs, e.g. 
\cite{Wuensche1997,AldanaEtAl2003}. One can measure, for example, the number
and length of attractors, the sizes and distributions of their basins, and
how these depend on different parameters of RBNs, such as $N$, $K$, $p$, or
the topology.

The structure of the nodes is very important for the dynamics of RBNs. The 
\emph{descendants} of a node are the nodes that it affects, while the \emph{%
ancestors} of a node are those that affect it. To have cycle attractors,
i.e. of period greater than one, there should be at least one node that will
be its own ancestor. A circuit of auto-activating nodes is called a \emph{%
linkage loop}, and when there is no feedback, \emph{linkage trees} are
formed. Note that loops spread activation through trees, but not vice versa.
The \emph{relevant} elements of a network are those nodes that form linkage
loops, and do not have constant functions, for these cause instabilities in
the network, which might or not propagate. Note that as there are more
connections in a network (higher $K$), the probability of having loops
increases. Therefore, finding less stable dynamics for high values of $K$ is
natural.

We should not confuse the node diagrams with the state space diagrams. These
show the dynamic trajectories of states of the network, while the first show
the relations of the network elements. Classic RBNs are dissipative
systems: a state can have only one \emph{successor}, since the dynamics are
deterministic, but more than one \emph{predecessor} or \emph{pre-image} can
flow into a single state. The \emph{in-degree} of a state is the number of
predecessors it has. States without predecessors are called \emph{%
garden-of-Eden} states. In this way, the dynamics flow from garden-of-Eden
states, converging towards attractors. The time it takes to reach an
attractor is called \emph{transient} time.

\subsubsection{Attractor Lengths}

There have been analytic solutions of RBNs for $K=1$ \cite
{FlyvbjergKjaer1988}, and for $K=N$ \cite{DerridaFlyvbjerg1987}. The
challenging problem of finding a general analytic solution is still open.
Some statistical studies have matched the analytic solutions for the special
cases. People have observed the following, considering $p=0.5$ \cite
{Kauffman1993,BastollaParisi1998,AldanaEtAl2003}:

For $K=1$, the probability of having long attractors decreases
exponentially, and the average number of cycles seems to be independent of $%
N $ \cite{BastollaParisi1998}. The median lengths of state cycles are of
order $\sqrt{N/2}$.

For $K\geq N$, the average length of attractors and the transient times
required to reach them grow exponentially. This restricts numerical
investigations to small networks. The typical cycle length grows
proportional to 2$^{N/2}$.

For $K=2$, at the critical phase, both the typical attractor lengths and the
average number of attractors grow algebraically with $N$. However, the
precise dependence of $N$ is a matter of dispute. People long believed that
the average number of attractors and their length was proportional to $\sqrt{N}$, \cite
{Kauffman1969,Kauffman1993,BastollaParisi1998}. This result was very attractive,
because the number of cell types and cell replication times for different organisms seem to scale also as the square root of
genes for different species, although the precise number of genes of
organisms keeps on changing. However, Bilke and Sjunnesson did a full
exploration of networks, ``decimating'' irrelevant variables, and found that
there is a \emph{linear} dependence of number of attractors depending on $N$ 
\cite{BilkeSjunnesson2002}. This linear dependence has been confirmed in
other complete statistical studies \cite{Gershenson2002e,GershensonEtAl2003a}%
. The difference seems to lie on the bias caused by \emph{undersampling} the
state space.

Since there are 2$^{N}$ states, full statistical studies are possible only
for very small networks. For example, for $N=20$ there are more than a
billion initial states. Therefore, these studies either concentrate on small
networks, or take into account only very few initial states. In the first
case, some properties of large networks will not be observed, whereas in the
second, some attractors would not be found, especially if their basins
consist of very few states.

More research is needed in this direction. One argument could be that even
when there would be potentially more attractors than the ones found by
limited sampling, in practice one would obtain the same result, since nature
does not exhaust all possible configurations (e.g. there could be more cell
types for a given number of genes, but \emph{developing} into them is
impossible, i.e. their basins are very small). Also, if we could get the
general analytical solution, it could be that it would not match statistical
studies, since a bias should be expected in the statistical sampling, due to
very high standard deviations. However, for different purposes we might be
more interested in the practical than the theoretical results, or vice versa.

\subsubsection{Convergence}

One can measure the convergence of states with different parameters. One of
them is the $G$-density, which is the density of garden-of-Eden states.
Another is the in-degree frequency distribution, which can be plotted as a
histogram \cite{Wuensche1998}. These measures reveal features at different
phases.

At the \textbf{ordered} phase, there is a very high $G$-density, and high
in-degree frequency. This leads to a high convergence, and very short
transient times. The basins of attraction are very compact, with many states
flowing into few states.

At the \textbf{critical} phase, the in-degree distribution approximates a
power-law, i.e. there are few states with high in-degree, and many states
with low in-degree. There is medium convergence.

At the \textbf{chaotic} phase, there are a relatively lower $G$-density, and
a high frequency of low in-degrees. The basins of attraction are very
elongated, with few states flowing into other states. This makes average
transient times very long, and in some cases infinite in practice.
Therefore, there is low convergence.

Other parameters that can be useful for measuring convergence include
Walker's ``internal homogeneity'' \cite{WalkerAshby1966}, Langton's $\lambda 
$\ parameter \cite{Langton1990}, and Wuensche's Z parameter \cite
{Wuensche1999}. The latter one, together with the ``input-entropy
variance'', can be also used to automatically classify rules of CA into
ordered, complex (critical), and chaotic \cite{Wuensche1999}. This is
useful, since ``interesting'' behaviour in CA tends to occur within complex
rule space.

\subsection{Multi-Valued Networks}

There have been some extensions to the Boolean idealization of classical
RBNs, namely where nodes can take more than two values.

Sol\'{e}, Luque, and Kauffman, and more recently Luque and Ballesteros have
studied such multi-valued networks, and calculated their phase transitions 
\cite{SoleEtAl2000,LuqueBallesteros2004}. For the special case where only
two states are allowed, the results of Derrida are recovered.

In nature, the components of certain systems exhibit a behaviour that is better
described with more than two states. Particular models should go beyond the
boolean idealization. However, for theoretical purposes, we could combine
several boolean nodes to act as a multi-valued one (codifying in base two
its state).

\subsection{Topologies}

Many systems have been found to have a scale-free topology. It seems to be a
persistent feature of complex networks \cite{Barabasi2002}: The Internet,
molecular and genetic networks, social networks, technology graphs, language
networks, food webs... they all share similar topological features: they
have few elements with many links, and many elements with few links. This
distribution seems to have several adaptive advantages, thought it is still
not very well understood.

However, most RBNs which have been studied have homogeneous or normal
topologies. Oosawa and Savageau studied the effects of topology in the
properties of RBNs \cite{OosawaSavageau2002}. Their results showed that the
topology can change drastically these properties. Networks with the more
uniform rank distributions exhibit more and longer attractors and less
entropy and mutual information (less correlation in their expression
patterns), whereas more skewed topologies exhibit less and shorter
attractors and more entropy and mutual information. A topology based on E.
coli, which is scale-free, balances the parameters to avoid the
disadvantages of the extreme topologies.

Having this in mind, Aldana studied many properties of RBNs with scale-free
topology \cite{Aldana2003}. The connectivity of scale-free RBNs can be
generated randomly using the probability distribution $P(k)=\left[ \zeta
\left( \gamma \right) k^{\gamma }\right] ^{-1}$, where $\gamma >1$ and $%
\zeta \left( \gamma \right) =\sum {}_{k=1}^{\infty }k^{-\gamma }$ is the
Riemann Zeta function. In this way, every node has at least one connection,
but there are few ones with many connections. The properties of the network
are no longer determined by the average connectivity, but by the exponent $%
\gamma $.

Following Derrida's method, Aldana found that the critical value of the
exponent $\gamma _{c}$ where the phase transition from order to chaos occurs
is determined by the transcendental equation:

\begin{equation}
2p\left( 1-p\right) \frac{\zeta \left( \gamma _{c}-1\right) }{\zeta \left(
\gamma _{c}\right) }=1  \label{eqAldana}
\end{equation}

\begin{figure}[t]
\begin{center}
\includegraphics[width=3.2in]{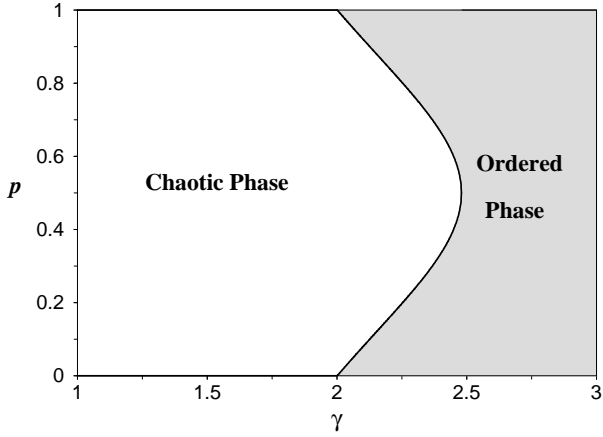}
\end{center}
\caption{Phase diagram for the scale-free model, reprinted from \protect\cite
{Aldana2003} with permission from Elsevier}
\label{Aldana-fig4}
\end{figure}

The values for which (\ref{eqAldana}) is satisfied are plotted in Figure \ref
{Aldana-fig4}. We can see that $\gamma _{c}\in \lbrack 2,2.5]$ for any value
of $p$. The maximum value of $\gamma _{c}\approx 2.47875$ is reached when $%
p=0.5$.

The network properties at each phase (e.g. number and length of attractors,
transient times) are analogous to the ones obtained with homogeneous RBNs.

An important result is that evolvability has more space in scale-free
networks, since these can adapt even in the ordered regime, where changes in
well-connected elements do propagate through the network. However,
experimental evidence shows that most biological networks are scale free
with exponent $2<\gamma <2.5$ \cite{Aldana2003}, i.e. ``at the edge of
chaos''. The advantages of scale-free topologies are beginning to become
evident, although they have been not embraced by most researchers of RBNs.

\subsection{RBN Control}

RBNs usually do not consider external inputs. However, real systems such as
genetic networks can be influenced by external signals, such as molecular
clocks related to sunlight.

Methods of chaos control have been successfully applied to chaotic RBNs \cite
{LuqueSole1997b,LuqueSole1998,BallesterosLuque2002}. The main idea is to use
a periodic function to drive a very chaotic network into a stable pattern.
If a periodic function determines the states of some nodes at some time,
these will have a regularity that can spread through the rest of the
network, developing into a global periodic pattern. A high percentage of
nodes should be controlled to achieve this. However, once we control a small
chaotic network, we can use this to control a larger chaotic network, and
this one to control an even larger one, and so on. This shows that it is possible to design
chaotic networks controlled by few external signals to force them into regular behaviour.

\section{Different Updating Schemes\label{sectUpdSchemes}}

Kauffman has argued that the small average number of attractors found in
RBNs compared with the number of possible states can account for the number
of cell types and cell replication time in organisms compared with their number of genes \cite
{Kauffman1969,Kauffman1993}. In those days, about eighty thousand genes were
thought to conform the human genome. Therefore, if the genome is seen as a
RBN close to the edge of chaos, the expected number of attractors would be
less than three hundred, matching the observed number of cell types in
humans. However, there are many drawbacks to this calculation. The Boolean
idealization has been roughly accepted, since multi-valued networks have
shown similar results. Now that the human genome has been mostly sequenced, it seems to consist of less than thirty
five thousand genes. The topology seems to be scale-free. There is certain
amount of junk or structural DNA, without functionality. Many functions seem
to be biassed ($p\neq 0.5$). But the heaviest argument has been the
following: \emph{genes do not march in step}. Genes do not change their
states all at the same moment, but some do it earlier than others. There was
no argument for the synchronicity in RBNs. In the next sections we will
review research made related to this criticism.

\subsection{Asynchronous RBNs}

Harvey and Bossomaier introduced the criticism to the synchronicity of
classic RBNs \cite{HarveyBossomaier1997}. It was well known that
asynchronicity could change drastically the dynamics of a synchronous
system, such as the prisoner's dilemma \cite{HubermanGlance1993} or Conway's
game of life \cite{BersiniDetours1994}, and they did a similar thing for
RBNs. Instead of updating the nodes synchronously, they defined asynchronous
RBNs (\textbf{ARBNs}), where a node is picked up at random, and updated. We
have to notice that ARBNs are not only asynchronous, but also
non-deterministic. This destroys the cycle attractors of classical RBNs (%
\textbf{CRBNs}), since it is very difficult that a sequence of states will
be repeated with a non-deterministic updating. Point attractors still appear
in ARBNs. There are also ``loose'' attractors, which can be seen as a subset
of the state space which ``traps'' the dynamics after some time (all the
possible states would rarely be visited).

\begin{figure}[t]
\begin{center}
\includegraphics[width=3.2in]{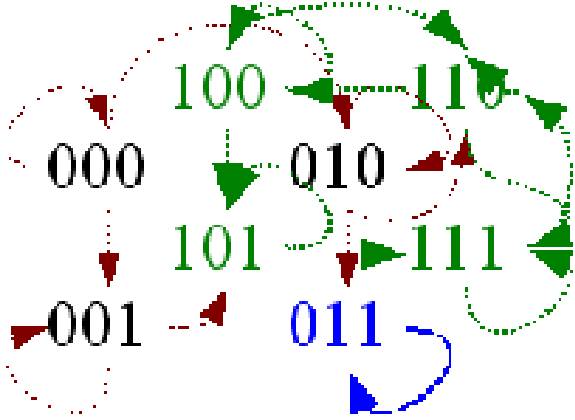}
\end{center}
\caption{State space diagram of an ARBN, using the lookup table and wiring
of Figure \ref{VgrCRBN}. Dotted arcs indicate stochastic transitions. There
is one point attractor (011) and one loose attractor (100,101,110,111). The state (0,0,0) could lead to either attractor, via (010) or (001), respectively; i.e. it is in the basin of both attractors}
\label{VgrARBN}
\end{figure}

The behaviour of ARBNs changes drastically from the one presented by CRBNs.
Not only cycle-attractors disappear, but their basins can change. Also,
states in ARBNs can be in more than one basin of attraction: they have the
potentiality to fall into different attractors depending on which nodes are
updated. An example of an ARBN can be appreciated in Figure \ref{VgrARBN}.

Harvey and Bossomaier found that in theory, there is on average exactly one
point attractor for any ARBN family. However, since we do not exhaust all
possible networks, a skewed distribution of the point attractors is made
evident: for some values ARBNs have more skewed distributions, i.e. few
networks with many point attractors, several with none, so the averages tend
to be lesser than one (except for the special case $K=0$, where it is
exactly one), reaching a minimum for $K=3$ \cite{Gershenson2002e}. It seems
that there are no loose attractors for $K=1$, and the probability of having
one increases with $K$.

The properties of ARBNs, being so different to the ones of RBNs, casted a
doubt on the validity of CRBNs as models of genetic regulatory networks.

\subsubsection{Rhythmic Asynchronous RBNs}

If ARBNs were to model biological systems, how could they have any rhythm?
An artificial way of solving this could be to implement a CRBN in an ARBN
using Nehaniv's method \cite{Nehaniv2002}, but such a network would be very
unrealistic. Another option could be to introduce proportionally large time
delays in each node, so that each node could be updated only when most
probably all the other nodes would have been updated \cite
{KlemmBornholdt2003}. However, this is in a way a disguised synchronicity, and
not more realistic than it.

Trying to find a solution, Di Paolo used genetic algorithms to explore the
space of possible ARBNs, and found networks that do have rhythmic behaviour 
\cite{DiPaolo2001}. Recently, Rholfshagen and Di Paolo analysed the topology
of such rhythmic ARBNs \cite{RohlfshagenDiPaolo2004}. They found that
invariably this consists of a ``ring'' of nodes that affect only one to the
next, i.e. a linkage loop. The rest of the nodes affect nodes only outside
the ring, so that activation can spread only outwards of the ring. In other
words, there is a single linkage loop in the network, and the dynamics of
this propagate only towards linkage trees. The number of nodes in the ring
determines the average period of the rhythm in epochs (time steps$\times N$%
). This is because only one node can change the guiding dynamics of the
network at a time, and any node is updated on average once an epoch. These
results are very promising, but there are open tasks, such as the
exploration of rhythmic ARBNs with more than one rhythmic attractor.

\subsection{Deterministic Asynchronous RBNs}

I agree with the criticisms to the synchronous assumption: genes do not
march in step. But I do not believe that they are random. Having this in
mind, I proposed Deterministic Asynchronous RBNs (\textbf{DARBNs}) \cite
{Gershenson2002e}.

For having nodes that do not update simultaneously, we can introduce two
parameters per node, $P_{i}$ and $Q_{i}$ ($P_{i},Q_{i}\in \mathbb{N}%
,P_{i}>Q_{i}$). All $P_{i}$'s and $Q_{i}$'s are generated randomly with maxima $P_{max}$ and $Q_{max}$, and remain fixed. A node $i$ will be updated when the modulus of the
time step over $P_{i}$ equals $Q_{i}$. Like this, $P_{i}$ can be seen as the 
\emph{period} of the node update, i.e. the number of time steps that will
pass between updated, and $Q_{i}$ can be seen as the \emph{translation }of
the node update. If at a certain time step more than one node fulfills its
updating condition, then the nodes will be updated in an arbitrary order
(e.g. from left to right), and the whole network will be updated with each
node.

We can also define Deterministic Generalized Asynchronous RBNs (\textbf{%
DGARBNs}), where if more than one node fulfills its updating condition, all
of these nodes will be updated synchronously. This makes DGARBNs
semi-synchronous.

\begin{figure*}[t]
\begin{center}
\includegraphics[width=6.5in]{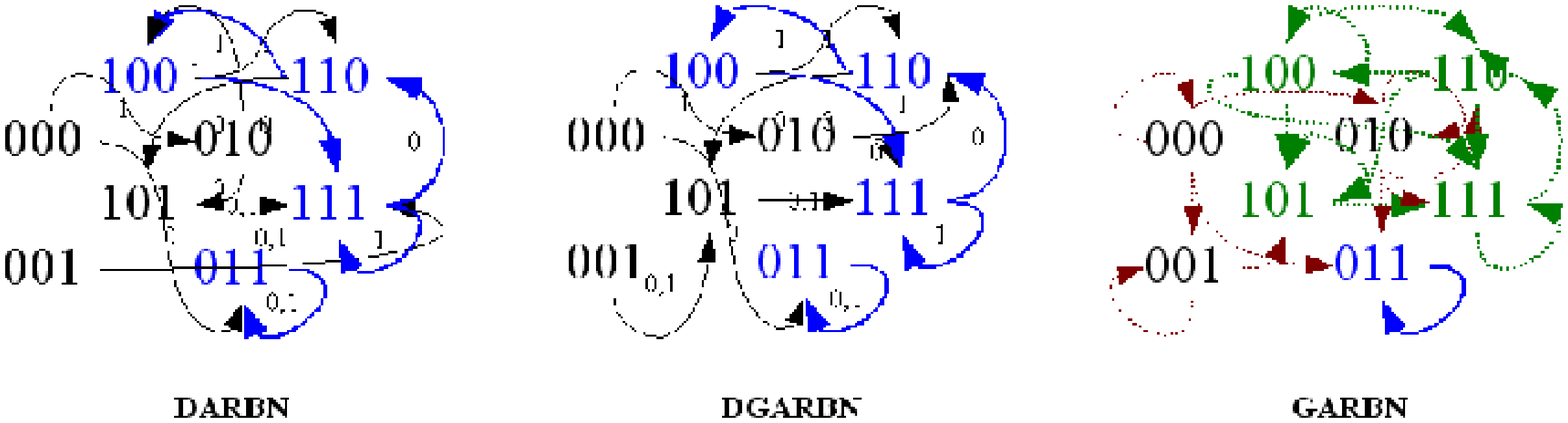}
\end{center}
\caption{State space diagrams of a DARBN, a DGARBN, and a GARBN, using the
lookup table and wiring of Figure \ref{VgrCRBN}. For the deterministic
networks, $\overline{P}=\left\{ 1,1,2\right\} $ and $\overline{Q}=\left\{
0,0,0\right\} $. Numbers near arcs indicate the transitions which will occur
at time modulus two. There is one point attractor (011) and one cycle
attractor, but different than the one of a CRBN ($100_{1}\rightarrow 111_{0}%
\rightarrow 111_{1}\rightarrow 110_{0}$), but the basins are different. For the GARBN, there is the same point
attractor and loose attractor than for the ARBN, but there are more possible
stochastic transitions, i.e. changes can potentially propagate faster}
\label{VgrDARBNs}
\end{figure*}

For completeness, we also defined Generalized Asynchronous RBNs (\textbf{%
GARBNs}), which are like ARBNs, only that at each time step, some nodes are
randomly selected, and these are updated synchronously. In this way, GARBNs
are semi-synchronous, but non-deterministic. We can see examples of these
RBNs in Figure \ref{VgrDARBNs}.

We found out that all types of networks have the same point attractors \cite
{Gershenson2002e}. This is because no matter which node is updated, the
state will not change. Therefore, the updating scheme does not affect point
attractors. However, the attractor basins do change, and other types of
attractor. DARBNs and DGARBNs have properties much closer to the ones of
CRBNs \cite{Gershenson2002e}. Since they are deterministic, cycle attractors
are present. The number of attractors, like in CRBNs, increases linearly
with $N$, but more slowly, meaning that there are even fewer attractors for
a high possible number of states\footnote{%
ARBNs and GARBNs have also a linear increase in the average number of
attractors, when we consider loose attractors in the statistics \cite
{Gershenson2004b}. These RBNs have less attractors than deterministic RBNs,
but their sizes are much larger.}. Also, the percentage of states in
attractors is reduced exponentially, as with CRBNs, but even faster. These
results imply that DGARBNs and DARBNs can perform even more \emph{complexity
reduction} than CRBNs. We also concluded that the difference of CRBNs and
ARBNs lies more in non-determinism than in asynchronicity.

Moreover, we proposed a method for mapping any deterministic asynchronous
RBN into a CRBN. The main idea is to introduce new nodes, connected to every
other previous node, which codify in base two the maximum period of the
DARBN. This is the least common multiple of all $P_{i}$'s.

\subsubsection{Asynchronicity and Feedback}

We have to mention that Thomas developed much earlier an asynchronous model
of RBNs using delays, which could be both deterministic or stochastic,
depending on the certainty of the delays \cite
{Thomas1973,Thomas1978,Thomas1991}. However, these models were proposed and
used mainly for the analysis of precise networks, their circuits, and
feedback loops, and to my knowledge they have been not used for analytical
or statistical studies of ensembles (``families'') of networks. An
interesting finding of Thomas and coworkers was the following: a positive
feedback loop (direct or indirect autocatalysis) in a network implies the
choice of two stable states, i.e. point attractors. This gives the property
of \emph{multistationarity}. On the other hand, a negative feedback loop
implies periodic behaviour, i.e. point or cycle attractors. This can be
described as \emph{homeostasis}. The combinations of positive and negative
feedback loops give RBNs a plethora of possible behaviours, many of which
are found in living systems.

\subsection{Mixed-context RBNs}

We can see the sets $\overline{P}$ and $\overline{Q}$, consisting of all $%
P_{i}$'s and $Q_{i}$'s respectively, as the \emph{context} of a network.
This is because external factors, such as temperature, can change the
updating periods of elements of a system. Like this, the same DGARBN can
have different behaviours in different contexts, i.e. for different $%
\overline{P}$ and $\overline{Q}$.

\begin{figure}[t]
\begin{center}
\includegraphics[width=3.2in]{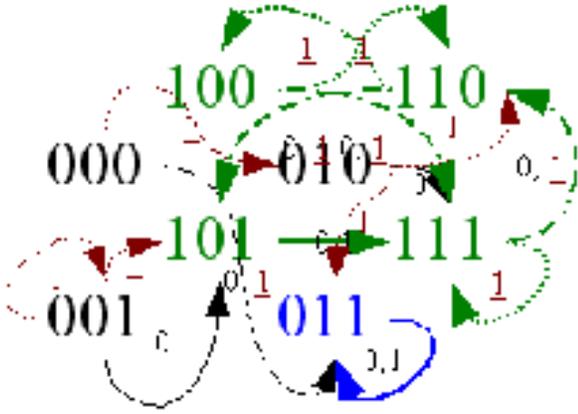}
\end{center}
\caption{State space diagram of a MxRBN, using the lookup table and wiring
of Figure \ref{VgrCRBN}. $M=2$ contexts. $\overline{P}_{1}=\left\{ 1,1,2\right\}
$,$\overline{P}_{2}=\left\{ 2,1,1\right\} $ and $\overline{Q}_{1}=\overline{Q}%
_{2}=\left\{ 0,0,0\right\} $. Therefore, the dynamics are deterministic and
context independent when the modulus of time over two is zero, and depend on
the context when it is one. There is one point attractor (011) and one loose
attractor (100,101,110,111)}
\label{VgrMxRBN}
\end{figure}

Having this in mind, we can introduce non-determinism in contextual RBNs in
a very specific way \cite{GershensonEtAl2003a}. For a given DGARBN, we can
have $M$ ``pure'' contexts. Then, each $R$ time steps we select randomly one
of these contexts, and use it in the network. Thus, we have defined
Mixed-context RBNs (\textbf{MxRBNs}). An example is shown in Figure \ref
{VgrMxRBN}.

We should note that the non-determinism in MxRBNs is introduced in a very
controlled fashion. In GARBNs we have $N$ ``coin flips'' per time step
(selecting which nodes will be updated). In ARBNs we have one coin flip per
time step (selecting which node will be updated). In MxRBNs we have one coin
flip per $R$ time steps (selecting which context will be used). The higher
the value of $R$ and the lower number of $M$ contexts, the less
stochasticity there will be.

MxRBNs have very similar number of attractors than ARBNs and GARBNs
(considering loose attractors), and also their number increases linearly and
slowly with $N$. However, they can perform significantly more \emph{%
complexity reduction} \cite{Gershenson2004b}, even more than CRBNs, but not
as much as DGARBNs or DARBNs. This means that very few states, from all
their possible states, lie in their attractors.

We can see that the \emph{context} in RBNs allows a high increase in
complexity reduction, since it allows information to be ``thrown'' into the
context. We can restate this with the following argument. Any updating scheme can perform in theory any
computation. The more non-determinism we introduce, the more redundancy in
the network we need to perform the computation. We require more elements. On
the other hand, \emph{contextual} RBNs can exploit information in their
contexts, which CRBNs need to codify explicitly. In terms of number of
elements required for a computation, less will be required in a DGARBN. Then
the ascending order would be of DARBNs, MxRBNs, CRBNs, ARBNs, and finally
GARBNs.

\subsection{Classification of RBNs}

\begin{figure}[t]
\begin{center}
\includegraphics[width=3.2in]{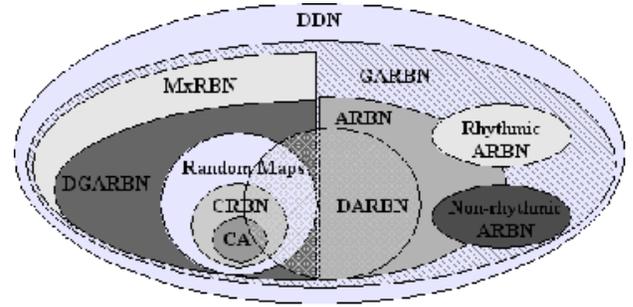}
\end{center}
\caption{Classification of random Boolean networks, according to their
updating scheme}
\label{RBNClassification}
\end{figure}

We can classify different types of RBNs according to their updating scheme 
\cite{Gershenson2002e}. All RBNs are discrete dynamical networks (DDNs) \cite
{Wuensche1997}, since they have discrete time, states, and values. The most
general type of RBNs are GARBNs, since all of the others can be seen as
special cases of them. If, on one hand, we make them deterministic with a
context conformed by $\overline{P}$ and $\overline{Q}$, then we would have
DGARBNs. MxRBNs would be more general than DGARBNs, since they have $M$
contexts. DGARBNs are a special case, when $M=1$ or $R\rightarrow \infty $.
Random maps would be a special case, of DGARBNs when $\overline{P}=\overline{%
1}$, $\overline{Q}=\overline{0}$, and $N=K$. Random maps can simulate with
redundancy any CRBN, but not vice versa, so the latter can be seen as a
special case of the former. Boolean CA are special cases of CRBNs, where the
connectivity and functionality are the same, i.e. symmetrical, for all
nodes. From GARBNs, on the other hand, we can limit to the update of only
one node at a time, and we will have ARBNs \cite{HarveyBossomaier1997}. If
we make them deterministic with a context, then we have DARBNs. These and DGARBNs overlap on the special cases
when one and only one node is updated at a time step. There are special
ARBNs with rhythmic and non-rhythmic attractors \cite{DiPaolo2001}, and
probably there are also GARBNs with these types of attractors. We should
note that particular instantiations of Thomas' asynchronous RBNs \cite
{Thomas1973,Thomas1978,Thomas1991} could be seen as ARBNs or DARBNs,
depending on the certainty of the delays. We can see a diagram of the
proposed classification in Figure \ref{RBNClassification}.

We could add a third dimension to Figure \ref{RBNClassification}, indicating
the number of allowed states per node, to include multi-valued networks.
Special cases of the different RBNs presented can be also interesting, such
as networks with scale-free or another particular topology, or CA with
different updating schemes. We can see that DDNs offer a rich variety of
models, many of which have not been yet studied.

\begin{figure*}[t]
\begin{center}
\includegraphics[width=6.5in]{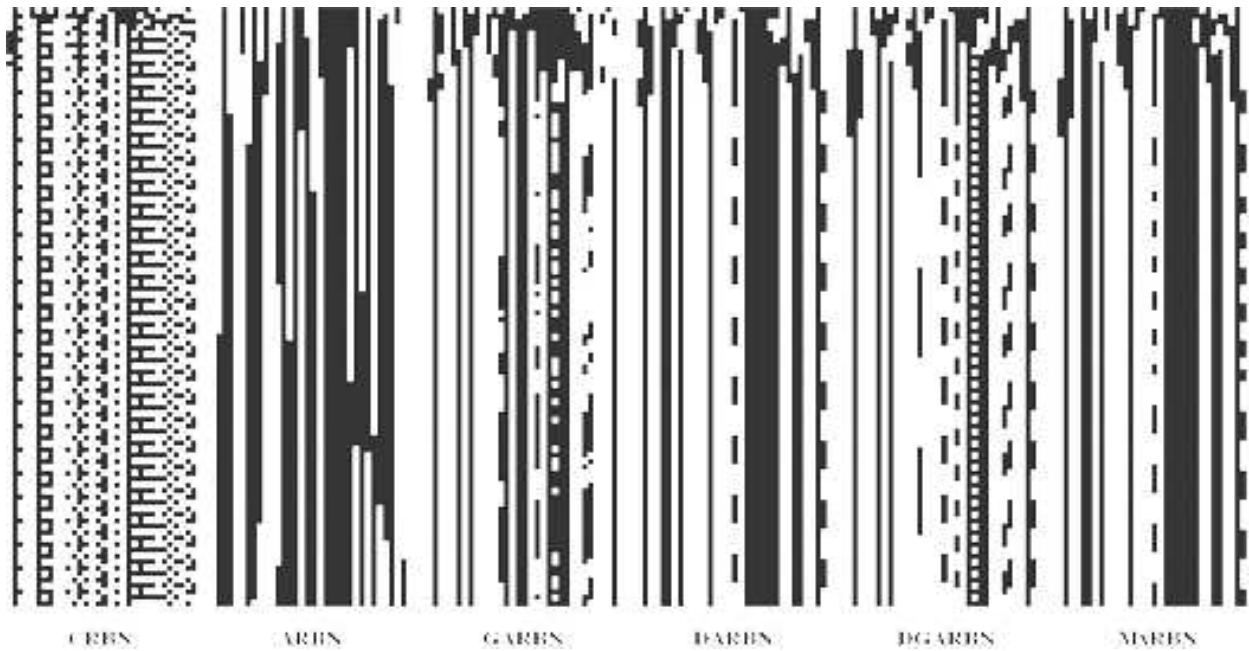}
\end{center}
\caption{Dynamics of the same RBN under different updating schemes, with $N=32, K=2, p=0.5$, for the same initial state. For contextual RBNs, $maxP=5, maxQ=4$. For MxRBN, $M=2, R=10$. Few states are frozen (e.g. second from left to right is always black, the fourth is always white), but dynamics change drastically with updating scheme. Note that in spite of being non-deterministic, the dynamics of the MxRBN are visually more similar to the ones of DARBN or DGARBN than ARBN or GARBN. 
}
\label{RBNsDynExample}
\end{figure*}

The more general a type of RBN is, the more trajectories in state space it
can have. Note that since we only change the updating scheme, we can easily
compare the behaviour of the same network, i.e. rules and connectivity, with
different schemes. We have seen that the dynamics and the basins of
attraction can change considerably as we switch updating scheme, although in
extreme cases, such as $K=0$ or when the number of attractors equals $N$,
different network types do have the same behaviour. An example can be seen in  Figure \ref{RBNsDynExample}. However, we have
recently found that, even when the precise stability of RBNs depends on the
updating scheme, the phase transition seems to be very similar independently
of the updating scheme \cite{Gershenson2004a}.

\section{Applications}

 The generality of RBNs makes them applicable in a wide range of
domains. In this section we review some of the main areas where RBNs have
been applied.

\subsection{Genetic Regulatory Networks}

Most cells in multicellular organisms have the same genetic information,
although the genes that are expressed or ``on'' change at every moment.
Genes interact with each other via proteins, but there are so many of them,
that genetic regulation is not fully understood. RBNs were originally
proposed to tackle this problem \cite{Kauffman1969}. They have been used not
only to explore their generic properties, but also to analyse and predict
genomic interactions \cite
{SomogyiSniegoski1996,SomogyiEtAl1997,D'haeseleer1998}. Since the genomic
data is incomplete, and certain knowledge about real genetic networks is
required for disease treatment, probabilistic boolean networks (\textbf{PBNs}%
) were introduced by Shmulevich and coworkers. These are useful for
inferring possible gene functionality from incomplete data \cite
{ShmulevichEtAl2002}.

There has been experimental evidence for Kauffman's interpretation of cell
types as attractors of RBNs \cite{HuangIngber2000}, which encourages more
RBN research. It seems that many genes do not play a role in the type of a
cell. They certainly might have other functions, for example related to the
cell's metabolism. Nevertheless, there is a very strong correlation for some
genes as a cell type is mechanically forced, which shows that the activation
patterns of a subset of genes can indicate the type of a cell.

There have been also other models of genetic regulatory networks which
include continuos states \cite{GlassKauffman1973,KapplerEtAl2002}. Using
differential equations in which gene interactions are incorporated as
logical functions, there is no need for a clock to calculate the dynamics.

\subsection{Evolution and Computation}

RBNs are very tempting models for studying evolution, since we do not assume
any functionality. It is now known that evolvability is expected at the edge
of chaos, where small changes do not destroy previous functionality, but can
explore their space of possibilities incrementally. RBNs also present
naturally modules, a desired feature in evolvable systems.

Fern\'{a}ndez and Sol\'{e} have studied issues related to the evolvability
of networks, such as robustness, redundancy, degeneracy, and modularity \cite
{FernandezSole2003}. This is a very abstract study of the requirements of
life, since one way we can distinguish physical from biological systems is
to note that the latter perform computations \cite{Hopfield1994}. By
understanding how a network can evolve to perform certain computations, we
are answering questions on how complex organisms evolved on Earth. There
have been some studies of evolution of RBNs using genetic algorithms with
promising results \cite{Stern1999,LemkeEtAl2001}.

Related work is the one carried out in the area of evolvable hardware, where
complex logical circuits are evolved in reconfigurable hardware \cite
{Thompson1998}. Also, research in RBNs could provide valuable feedback for
evolving logical circuits.

\subsection{Other areas}

RBNs have been used in several other areas, such as neural networks \cite
{HuepeAldana2002}, social modelling \cite{Shelling1971}, robotics \cite{QuickEtAl2003}, and music
generation \cite{Dorin2000}.

Finally, RBNs are interesting mathematical objects by themselves. Since
cellular automata are special cases of RBNs, we can find many more
applications there, e.g. in percolation theory \cite{Stauffer1985}.

\section{Tools}

There are several software applications available for the exploration of
different properties of RBNs:

\textbf{DDLab}. Developed by Andy Wuensche, Discrete Dynamics Lab is
probably the most powerful tool for studying discrete dynamical networks:
synchronous RBNs and CA, including multi-valued networks. One can observe
the dynamics of the networks, explore their basins of attraction. It
includes a wide variety of measures, data, analysis and statistics. It is
very well documented, and runs on most platforms. It is available at
http://www.ddlab.com

\textbf{RBN Toolbox} for Matlab was developed by Christian Schwarzer and
Christof Teuscher. It is useful for simulating and visualizing RBNs. It
includes different updating schemes, statistical functions, and other
features. It is available at http://www.teuscher.ch/rbntoolbox

\textbf{RBNLab} was developed by the author. It can graphically simulate the
dynamics of RBNs with different updating schemes. It can find attractors
(point, cycle, and loose), and generate different statistics. It is
implemented in Java, so it runs on most platforms. The source code and the
program are available at http://rbn.sourceforge.net

\textbf{BN/PBN Toolbox} for Matlab is maintained by Harri L\"{a}hdesm\"{a}ki
and Ilya Shmulevich. It can be used to work with CRBNs and Probabilistic
Boolean Networks. It includes functions for simulating the network dynamics,
computing network statistics (numbers and sizes of attractors, basins,
transient lengths, Derrida curves, percolation on 2-D lattices, influence
matrices), computing state transition matrices and obtaining stationary
distributions, inferring networks from data, generating random networks and
functions, visualization and printing, intervention, and membership testing
of Boolean functions. It is available at
http://www2.mdanderson.org/app/ilya/PBN/PBN.htm

\section{Future Lines of Research}

There are many promising lines to be followed in RBN research. The ensemble
approach is promising for understanding and predicting properties of cells
and organisms \cite{Kauffman2004}. The use of RBNs for data mining and
genetic network analysis is being propagated \cite
{D'haeseleer1998,ShmulevichEtAl2002}. Because of their generality, RBNs are
also interesting for studying evolvability and adaptability at an abstract
level. Generalizations, combinations, and refinements of the different types
of RBNs presented are also worth pursuing, e.g. ensemble studies on Thomas'
ARBNs or PBNs, scale-free multi-valued DGARBNs, etc. Also, the challenging
problem of finding analytical solutions for CRBNs is still evading us. In
short, there is plenty of RBN research to do in the years to come.

\section{Conclusions}

This tutorial was a brief introduction to some of the research related to
random Boolean networks in the past decades. There has been much more work
that could not be included because of different constrains. However, the
readers are invited to deepen their knowledge in RBNs following the
references of this tutorial.

There will be many arguments on which will be the ``best'' model for different
purposes and phenomena. CRBNs may be not very close to reality, but full
ARBNs are even farther, since genes are not updated in a fully stochastic
manner\footnote{%
We have found out that ARBNs and GARBNs perform less complexity reduction
than deterministic RBNs or MxRBNs \cite{Gershenson2004a}. This indicates
that determinism or quasi-determinism is a desirable property of natural
systems. How could this evolve, is a different question, and studies such as
the ones on rhythmic ARBNs \cite{DiPaolo2001} are providing interesting
answers.}. Models such as DGARBNs or MxRBNs seem to be more realistic,
although they are more complicated. For ensemble studies, for simplicity,
CRBNs can be justifiable \cite{Gershenson2004b}. For modelling particular
genetic networks, other models might be more realistic and useful, such as
PBNs, or the models of Thomas, Glass, or Somogyi. However, their complexity
would make difficult, but still interesting, ensemble studies on them. We
can say that different models will be more suitable for different purposes,
and most of them are worth exploring.

We can say that RBNs are very inviting, because of their generality: one can
model at a very abstract level many phenomena, and study generic properties
of networks independently of their functionality. They have also raised
important questions related to the differences between theory and practice,
since some analytical or statistical results do not match each other. This
can be explained with skewed distributions and very high standard deviations
found in RBN statistics. However, this brings deeper philosophical
questions: how much should we care about theory, and how much should we care
about practice? Which one will help us understand better the phenomena we
try to study? It seems that a careful \emph{balance} of both is required.
However, it is unknown how this balance might be like.

\section{Acknowledgements}

This work was enriched with discussions, suggestions, and assistance from
Diederik Aerts, Maximino Aldana, Jan Broekaert, Keith Campbell, Ezequiel Di Paolo, Inman
Harvey, Bernardo Huberman, Stuart Kauffman, Ilya Shmulevich, and Andy
Wuensche. I thank Marcelle Kaufman for introducing me to the work of
Ren\'{e} Thomas. This work was supported in part by the Consejo Nacional de
Ciencia y Tecnolog\'{i}a (CONACYT) of M\'{e}xico.

\bibliographystyle{alife9}
\bibliography{carlos,RBN}

\begin{thebibliography}{}

\bibitem[Aldana, 2003]{Aldana2003}
Aldana, M. (2003).
\newblock Boolean dynamics of networks with scale-free topology.

\bibitem[Aldana-Gonz{\'a}lez et~al., 2003]{AldanaEtAl2003}
Aldana-Gonz{\'a}lez, M., Coppersmith, S., and Kadanoff, L.~P. (2003).
\newblock Boolean dynamics with random couplings.
\newblock In Kaplan, E., Marsden, J.~E., and Sreenivasan, K.~R., editors, {\em
  Perspectives and Problems in Nonlinear Science. A Celebratory Volume in Honor
  of Lawrence Sirovich}. Springer Applied Mathematical Sciences Series.

\bibitem[Ballesteros and Luque, 2002]{BallesterosLuque2002}
Ballesteros, F.~J. and Luque, B. (2002).
\newblock Random {Boolean} networks response to external periodic signals.
\newblock {\em Physica A}, 313:289 – 300.

\bibitem[Barab{\'a}si, 2002]{Barabasi2002}
Barab{\'a}si, A.-L. (2002).
\newblock {\em Linked: The New Science of Networks}.
\newblock Perseus.

\bibitem[Bastolla and Parisi, 1998]{BastollaParisi1998}
Bastolla, U. and Parisi, G. (1998).
\newblock Relevant elements, magnetization and dynamical properties in
  {Kauffman} networks: A numerical study.
\newblock {\em Physica D}, 115(3'4):203--218.

\bibitem[Bersini and Detours, 1994]{BersiniDetours1994}
Bersini, H. and Detours, V. (1994).
\newblock Asynchrony induces stability in cellular automata based models.
\newblock In {\em Proceedings of the {IVth} Conference on Artificial Life},
  pages 382--387. MIT Press.

\bibitem[Bilke and Sjunnesson, 2002]{BilkeSjunnesson2002}
Bilke, S. and Sjunnesson, F. (2002).
\newblock Stability of the {Kauffman} model.
\newblock {\em Physical Review E}, 65(016129).

\bibitem[Derrida and Flyvbjerg, 1987]{DerridaFlyvbjerg1987}
Derrida, B. and Flyvbjerg, H. (1987).
\newblock The random map model: A disordered model with deterministic dynamics.
\newblock {\em J. Physique}, 48:971–978.

\bibitem[Derrida and Pomeau, 1986]{DerridaPomeau1986}
Derrida, B. and Pomeau, Y. (1986).
\newblock Random networks of automata: A simple annealed approximation.
\newblock {\em Europhys. Lett.}, 1(2):45--49.

\bibitem[D'haeseleer et~al., 1998]{D'haeseleer1998}
D'haeseleer, P., Wen, X., Fuhrman, S., and Somogyi, R. (1998).
\newblock Mining the gene expression matrix: Inferring gene relationships from
  large scale gene expression data.
\newblock In Paton, R.~C. and Holcombe, M., editors, {\em Information
  Processing in Cells and Tissues}, pages 203--212. Plenum Publishing.

\bibitem[{Di Paolo}, 2001]{DiPaolo2001}
{Di Paolo}, E.~A. (2001).
\newblock Rhythmic and non-rhythmic attractors in asynchronous random {Boolean}
  networks.
\newblock {\em Biosystems}, 59(3):185--195.

\bibitem[Dorin, 2000]{Dorin2000}
Dorin, A. (2000).
\newblock Boolean networks for the generation of rhythmic structure.
\newblock In Brown and Wilding, editors, {\em Proceedings Australian Computer
  Music Conference 2000}, pages 38--45.

\bibitem[Fern{\'a}ndez and Sol{\'e}, 2004]{FernandezSole2003}
Fern{\'a}ndez, P. and Sol{\'e}, R. (2004).
\newblock The role of computation in complex regulatory networks.
\newblock In Koonin, E.~V., Wolf, Y.~I., and Karev, G.~P., editors, {\em Power
  Laws, Scale-Free Networks and Genome Biology}, number 03-10-055. Landes
  Bioscience.

\bibitem[Flyvbjerg and Kjaer, 1988]{FlyvbjergKjaer1988}
Flyvbjerg, H. and Kjaer, N. (1988).
\newblock Exact solution of the {Kauffman} model with connectivity one.
\newblock {\em J. Phys. A}, 21(7):1695–1718.

\bibitem[Gershenson, 2002]{Gershenson2002e}
Gershenson, C. (2002).
\newblock Classification of random {Boolean} networks.
\newblock In Standish, R.~K., Bedau, M.~A., and Abbass, H.~A., editors, {\em
  Artificial Life {VIII}: Proceedings of the Eight International Conference on
  Artificial Life}, pages 1--8. MIT Press.

\bibitem[Gershenson, 2004a]{Gershenson2004a}
Gershenson, C. (2004a).
\newblock Phase transitions in random {Boolean} networks with different
  updating schemes.
\newblock Submitted to Physica D.

\bibitem[Gershenson, 2004b]{Gershenson2004b}
Gershenson, C. (2004b).
\newblock Updating schemes in random {Boolean} networks: Do they really matter?
\newblock In {\em Artificial Life {IX}}. MIT Press.

\bibitem[Gershenson et~al., 2003]{GershensonEtAl2003a}
Gershenson, C., Broekaert, J., and Aerts, D. (2003).
\newblock Contextual random {Boolean} networks.
\newblock In Banzhaf, W., Christaller, T., Dittrich, P., Kim, J.~T., and
  Ziegler, J., editors, {\em Advances in Artificial Life, 7th European
  Conference, {ECAL} 2003 {LNAI} 2801}, pages 615--624. Springer-Verlag.

\bibitem[Glass and Kauffman, 1973]{GlassKauffman1973}
Glass, L. and Kauffman, S. (1973).
\newblock The logical analysis of continuous, nonlinear biochemical control
  networks.
\newblock {\em Journal of Theoretical Biology}, 39:103--129.

\bibitem[Harvey and Bossomaier, 1997]{HarveyBossomaier1997}
Harvey, I. and Bossomaier, T. (1997).
\newblock Time out of joint: Attractors in asynchronous random {Boolean}
  networks.
\newblock In Husbands, P. and Harvey, I., editors, {\em Proceedings of the
  Fourth European Conference on Artificial Life {(ECAL97)}}, pages 67--75. MIT
  Press.

\bibitem[Hopfield, 1994]{Hopfield1994}
Hopfield, J.~J. (1994).
\newblock Physics, computation, and why biology looks so different.
\newblock {\em Journal of Theoretical Biology}, 171:53--60.

\bibitem[Huang and Ingber, 2000]{HuangIngber2000}
Huang, S. and Ingber, D.~E. (2000).
\newblock Shape-dependent control of cell growth, differentiation, and
  apoptosis: Switching between attractors in cell regulatory networks.
\newblock {\em Exp. Cell Res.}, 261:91--103.

\bibitem[Huberman and Glance, 1993]{HubermanGlance1993}
Huberman, B.~A. and Glance, N.~S. (1993).
\newblock Evolutionary games and computer simulations.
\newblock {\em Proc. Natl. Acad. Sci. USA}, 90:7716--7718.

\bibitem[Huepe-Minoletti and Aldana-Gonz{\'a}lez, 2002]{HuepeAldana2002}
Huepe-Minoletti, C. and Aldana-Gonz{\'a}lez, M. (2002).
\newblock Dynamical phase transition in a neural network model with noise: An
  exact solution.
\newblock {\em Journal of Statistical Physics}, 108(3/4):527--540.

\bibitem[Kappler et~al., 2002]{KapplerEtAl2002}
Kappler, K., Edwards, R., and Glass, L. (2002).
\newblock Dynamics in high dimensional model gene networks.
\newblock {\em Signal Processing}, 83:789--798.

\bibitem[Kauffman, 1969]{Kauffman1969}
Kauffman, S.~A. (1969).
\newblock Metabolic stability and epigenesis in randomly constructed genetic
  nets.
\newblock {\em Journal of Theoretical Biology}, 22:437--467.

\bibitem[Kauffman, 1993]{Kauffman1993}
Kauffman, S.~A. (1993).
\newblock {\em The Origins of Order}.
\newblock Oxford University Press.

\bibitem[Kauffman, 2000]{Kauffman2000}
Kauffman, S.~A. (2000).
\newblock {\em Investigations}.
\newblock Oxford University Press.

\bibitem[Kauffman, 2004]{Kauffman2004}
Kauffman, S.~A. (2004).
\newblock The ensemble approach to understand genetic regulatory networks.
\newblock {\em Physica A: Statistical Mechanics and its Applications}, In
  Press.

\bibitem[Klemm and Bornholdt, 2003]{KlemmBornholdt2003}
Klemm, K. and Bornholdt, S. (2003).
\newblock Robust gene regulation: Deterministic dynamics from asynchronous
  networks with delay.
\newblock q-bio/0309013.

\bibitem[Langton, 1990]{Langton1990}
Langton, C. (1990).
\newblock Computation at the edge of chaos: Phase transitions and emergent
  computation.
\newblock {\em Physica D}, 42:12--37.

\bibitem[Lemke et~al., 2001]{LemkeEtAl2001}
Lemke, N., Mombach, J. C.~M., and Bodmann, B. E.~J. (2001).
\newblock A numerical investigation of adaptation in populations of random
  {Boolean} networks.
\newblock {\em Physica A}, 301(1-4):589--600.

\bibitem[Luque and Ballesteros, 2004]{LuqueBallesteros2004}
Luque, B. and Ballesteros, F.~J. (2004).
\newblock Random walk networks.
\newblock {\em Physica A: Statistical Mechanics and its Applications}, In
  Press.

\bibitem[Luque and Sol{\'e}, 1997a]{LuqueSole1997b}
Luque, B. and Sol{\'e}, R.~V. (1997a).
\newblock Controlling chaos in {Kauffman} networks.
\newblock {\em Europhys. Lett.}, 37(9):597--602.

\bibitem[Luque and Sol{\'e}, 1997b]{LuqueSole1997}
Luque, B. and Sol{\'e}, R.~V. (1997b).
\newblock Phase transitions in random networks: Simple analytic determination
  of critical points.
\newblock {\em Physical Review E}, 55(1):257--260.

\bibitem[Luque and Sol{\'e}, 1998]{LuqueSole1998}
Luque, B. and Sol{\'e}, R.~V. (1998).
\newblock Stable core and chaos control in random {Boolean} networks.
\newblock {\em J. Phys. A: Math. Gen.}, 31:1533--1537.

\bibitem[Luque and Sol{\'e}, 2000]{LuqueSole2000}
Luque, B. and Sol{\'e}, R.~V. (2000).
\newblock Lyapunov exponents in random {Boolean} networks.
\newblock {\em Physica A}, 284:33--45.

\bibitem[Mitchell et~al., 1993]{MitchellEtAl1993}
Mitchell, M., Crutchfield, J.~P., and Hraber, P.~T. (1993).
\newblock Dynamics, computation, and the "edge of chaos": A re-examination.
\newblock Technical Report 93-06-040, Santa Fe Institute.

\bibitem[Nehaniv, 2002]{Nehaniv2002}
Nehaniv, C.~L. (2002).
\newblock Evolution in asynchronous cellular automata.
\newblock In Standish, R.~K., Bedau, M.~A., and Abbass, H.~A., editors, {\em
  Artificial Life {VIII}: Proceedings of the Eight International Conference on
  Artificial Life}, pages 75--73. MIT Press.

\bibitem[Oosawa and Savageau, 2002]{OosawaSavageau2002}
Oosawa, C. and Savageau, M.~A. (2002).
\newblock Effects of alternative connectivity on behavior of randomly
  constructed {Boolean} networks.
\newblock {\em Physica D}, 170:143--161.

\bibitem[Quick et~al., 2003]{QuickEtAl2003}
Quick, T., Nehaniv, C.~L., Dautenhahn, K., and Roberts, G. (2003).
\newblock Evolving embodied genetic regulatory network-driven control systems.
\newblock In Banzhaf, W., Christaller, T., Dittrich, P., Kim, J.~T., and
  Ziegler, J., editors, {\em Advances in Artificial Life: {ECAL} 2003}, pages
  266--277. Springer.

\bibitem[Rholfshagen and {Di Paolo}, 2004]{RohlfshagenDiPaolo2004}
Rholfshagen, P. and {Di Paolo}, E.~A. (2004).
\newblock The circular topology of rhythm in random asynchronous boolean
  networks.
\newblock {\em BioSystems}, 73(2):141--152.

\bibitem[Shelling, 1971]{Shelling1971}
Shelling, T.~C. (1971).
\newblock Dynamic models of segregation.
\newblock {\em Journal of Mathematical Sociology}, 1:143–186.

\bibitem[Shmulevich et~al., 2002]{ShmulevichEtAl2002}
Shmulevich, I., Dougherty, E., and Zhang, W. (2002).
\newblock From {Boolean} to probabilistic {Boolean} networks as models of
  genetic regulatory networks.
\newblock {\em Proceedings of the IEEE}, 90(11):1778--1792.

\bibitem[Sol{\'e} et~al., 2000]{SoleEtAl2000}
Sol{\'e}, R.~V., Luque, B., and Kauffman, S.~A. (2000).
\newblock Phase transitions in random networks with multiple states.
\newblock Technical Report 00-02-011, Santa Fe Institute.

\bibitem[Somogyi et~al., 1997]{SomogyiEtAl1997}
Somogyi, R., Fuhrman, S., Askenazi, M., and Wuensche, A. (1997).
\newblock The gene expression matrix: Towards the extraction of genetic network
  architectures.
\newblock {\em Nonlinear Analysis: Theory, Methods and Applications},
  30(3):1815--1824.

\bibitem[Somogyi and Sniegoski, 1996]{SomogyiSniegoski1996}
Somogyi, R. and Sniegoski, C.~A. (1996).
\newblock Modeling the complexity of genetic networks: Understanding multigenic
  and pleiotropic regulation.
\newblock {\em Complexity}, 1(6):45--63.

\bibitem[Stauffer, 1985]{Stauffer1985}
Stauffer, D. (1985).
\newblock {\em Introduction to Percolation Theory}.
\newblock Taylor and Francis, London.

\bibitem[Stern, 1999]{Stern1999}
Stern, M.~D. (1999).
\newblock Emergence of homeostasis and noise imprinting in an evolution model.
\newblock {\em PNAS}, 96:10746--10751.

\bibitem[Thomas, 1973]{Thomas1973}
Thomas, R. (1973).
\newblock Boolean formalization of genetic control circuits.
\newblock {\em J. Theor. Biol.}, 42:563--585.

\bibitem[Thomas, 1978]{Thomas1978}
Thomas, R. (1978).
\newblock Logical analysis of systems comprising feedback loops.
\newblock {\em J. Theor. Biol.}, 73:631--656.

\bibitem[Thomas, 1991]{Thomas1991}
Thomas, R. (1991).
\newblock Regulatory networks seen as asynchronous automata: A logical
  description.
\newblock {\em J. Theor. Biol.}, 153:1--23.

\bibitem[Thompson, 1998]{Thompson1998}
Thompson, A. (1998).
\newblock {\em Hardware Evolution: Automatic Design of Electronic Circuits in
  Reconfigurable Hardware by Artificial Evolution}.
\newblock Distinguished dissertation series. Springer-Verlag.

\bibitem[{von Neumann}, 1966]{vonNeumann1966}
{von Neumann}, J. (1966).
\newblock {\em The Theory of Self-Reproducing Automata}.
\newblock University of Illinois Press.

\bibitem[Walker and Ashby, 1966]{WalkerAshby1966}
Walker, C. and Ashby, W. (1966).
\newblock On the temporal characteristics of behavior in certain complex
  systems.
\newblock {\em Kybernetik}, 3(2):100--108.

\bibitem[Wolfram, 1986]{Wolfram1986}
Wolfram, S. (1986).
\newblock {\em Theory and Application of Cellular Automata}.
\newblock World Scientific.

\bibitem[Wuensche, 1997]{Wuensche1997}
Wuensche, A. (1997).
\newblock {\em Attractor Basins of Discrete Networks}.
\newblock PhD thesis, University of Sussex.

\bibitem[Wuensche, 1998]{Wuensche1998}
Wuensche, A. (1998).
\newblock Discrete dynamical networks and their attractor basins.
\newblock In Standish, R., Henry, B., Watt, S., Marks, R., Stocker, R., Green,
  D., Keen, S., and Bossomaier, T., editors, {\em Complex Systems '98}, pages
  3--21, University of New South Wales, Sydney, Australia.

\bibitem[Wuensche, 1999]{Wuensche1999}
Wuensche, A. (1999).
\newblock Classifying cellular automata automatically: Finding gliders,
  filtering, and relating space-time patterns, attractor basins, and the z
  parameter.
\newblock {\em Complexity}, 4(3):47--66.

\bibitem[Wuensche and Lesser, 1992]{WuenscheLesser1992}
Wuensche, A. and Lesser, M. (1992).
\newblock {\em The Global Dynamics of Cellular Automata; An Atlas of Basin of
  Attraction Fields of One-Dimensional Cellular Automata}.
\newblock Santa Fe Institute Studies in the Sciences of Complexity.
  Addison-Wesley, Reading, MA.

\end{thebibliography}

\end{document}